\documentstyle[prl,aps]{revtex}
\def\ea{{\it et al.}}

\font\nrm=cmr9

\begin{document}
\draft
\twocolumn[\hsize\textwidth\columnwidth\hsize\csname@twocolumnfalse%
\endcsname
\phantom{Reply to Bernevig, Giuliano, and Laughlin}
%\title{\phantom{Reply to Bernevig, Giuliano, and Laughlin}}
%\maketitle
]

\vspace{10pt}
\noindent {\bf Greiter and Schuricht reply:}\ \ \
In response to a recent manuscript of ours \cite{greiterschuricht04},
Bernevig, Giuliano, and Laughlin (BGL) have reiterated their claim
that spinons in the Haldane--Shastry model (HSM) are interacting in a
comment posted on cond-mat
\cite{BernevigGiulianoLaughlin04condmat}.  Notwithstanding that this
comment does not contain any truly new arguments, that is, no argument
which has not been presented in some form or variation in one of their
previous papers
\cite{BernevigGiulianoLaughlin01prl,BernevigGiulianoLaughlin01prb} and
subsequently been disputed in our manuscript
\cite{greiterschuricht04}, we feel a certain obligation to respond to
each point they make.  We should caution, however, that some readers
may perceive the discussion as circular.

To begin with, BGL acknowledge that the $S$-matrix for spinon-spinon
scattering in the HSM does not depend on the pseudomomenta used in the
framework of the asymptotic Bethe Ansatz (ABA) \cite{Essler95}, while
maintaining that there is an spinon-spinon interaction.  They claim
that ``the apparent contradiction [...]\ is just a consequence of a
different way of labeling spinons''.  In other words, they assert that
the spinon $S$-matrix does depend on the true and physical spinon
momenta, while it does not depend on the pseudomomenta (which they
refer to as ``quasimomenta''), which, according to them, ``are good
quantum numbers when exactly solving correlated Hamiltonians but
[...]\ are not observable''.

It is not very difficult to see that this line of thought is
unsustainable.  The pseudomomenta in the ABA solutions are the quantum
numbers which label the exact eigenstates of the Hamiltonian, and in
particular the spin polarized two-spinon eigenstates investigated by
BGL.  Regardless of the physical interpretation of these numbers, if
the spinon scattering matrix $S$ does not depend on them, it does not
depend on the quantum numbers labeling the states.  Hence $S$ cannot
depend on the true and physical spinon momenta either, whatever they
may be.  The scattering matrix is an unambiguously defined quantity.
In the case of spinons in the HSM, %$S=\pm i$ 
$S=i$ directly and unambivalently implies that the spinons are
non-interacting particles with half-fermi statistics.

BGL further emphasize that the decomposition of the basis states
$\Psi_{\alpha\beta}$ in terms of the energy eigenstates $\Phi_{mn}$ is
unambiguous, and that the latter basis does not suffer from
overcompleteness.  This is in no way contrary to our observations: it
is the basis $\Psi_{\alpha\beta}$ which is overcomplete, and this
renders an interpretation of $\alpha$ and $\beta$ as spinon
coordinates ambiguous.  It is hence neither possible to interpret
$p_{nm}(\eta_\alpha-\eta_\beta)$ as the relative wave function of the
two spinons, nor possible to interpret an enhancement in this
quantity for short separations as evidence for an attraction, as
explained at length in \cite{greiterschuricht04}.  
Haldane had a good reason to refer to $\alpha$ and $\beta$ as ``spinon
coordinates'' in quotation marks \cite{Haldane91prl1}.
 
Finally, BGL claim that the identification of the individual spinon
momenta we propose ((17) of \cite{greiterschuricht04}) is unphysical.
In this context, we wish to remark that it is only possible to read of
the total momentum of the state (7) (the equation numbers here and
below refer to \cite{greiterschuricht04}) that is, the sum of the
momenta of the two spinons,
\begin{displaymath}
  q_m+q_n=\pi-\frac{2\pi}{N}\left(m+n+1\right),
\end{displaymath}
where $m\ge n$.  BGL assign the individual spinon momenta according to
(16), while we propose (17).  Let us emphasize here that there is no
reason to assume that the functional form of both $q_m$ and $q_n$ as
functions of $m$ and $n$ (as $m\ge n$) should be the same.  Our
proposal (17) has no drawbacks in comparison (16), but the advantage
that the relative momentum spacing between the spinons take the values
appropriate for half-fermions, while the spacings of (16) would be
appropriate for bosons.  (It was established by Haldane that the
spinons are half-fermions \cite{Haldane91prl2}.)    Our expression
for the two-spinon energies (18) is consistent with the Bethe Ansatz
result, while the corresponding expression obtained with BGL's
proposal (16) would yield a spinon-spinon interaction term.  The ABA
solution of the model, however, precludes such an interaction, as
reemphasized above.  In our opinion, it is fair to conclude that (17)
is the physically correct assignment.  In any occasion, it is not
in the scientific tradition to base a claim on rather arbitrarily
assuming (16).

In the conclusion of their comment, BGL state that the ABA result,
which states that the spinons in the HSM are free, does not
contradict their conclusion that they are interacting, but rather 
complements it.  In fact, however, these statements do contradict
each other, and there are only two possible resolutions.  Either
the ABA is not applicable to the model, and therefore not able to 
provide reliable results, or it is applicable, and the spinons are
free.  We believe we have unambiguously dispersed all of BGL's 
evidence against the latter.  
% We have established that there is no interaction between spinons.
There is no attraction between spinons in the HSM.

This work was %partially 
supported in parts by the German Research Foundation (DFG) through GK 284.

\vspace{10pt}
\noindent\hbox{Martin Greiter and Dirk Schuricht\hfill}

\medskip
\vbox{%\baselineskip=12pt
\obeylines\nrm 
Institut f\"ur Theorie der Kondensierten Materie
Universit\"at Karlsruhe
Postfach 6980
D-76128 Karlsruhe}

\vspace{5pt}
\noindent\hbox{\nrm December 2, 2004\hfill}

\noindent\hbox{\nrm PACS numbers: 75.10.Pq, 02.30.Ik, 75.10.Jm, 75.50.Ee\hfill}

%\begin{thebibliography}{10}

%\end{thebibliography}

\end{document}